\documentstyle[12pt]{article}

 \newcommand{\be}{\begin{equation}}
 \newcommand{\ee}{\end{equation}}
 \newcommand{\ba}{\begin{eqnarray}}
 \newcommand{\ea}{\end{eqnarray}}
 
 \newcommand{\del}{\partial}

\newcommand{\lef}{\left}

\newcommand{\ri}{\right}

\newcommand{\ca}{{\cal A}}

\newcommand{\fr}{\frac}

\begin{document}

\begin{titlepage}

\topmargin -15mm

\vskip 10mm

\centerline{ \LARGE\bf Phase Structure and Universality }
\vskip 3mm
\centerline{ \LARGE\bf  in Two-Dimensional Disordered }
\vskip 3mm
\centerline{ \LARGE\bf Quantum Antiferromagnets }

\vskip 2.0cm

\centerline{\sc E.C.Marino}

\vskip 0.6cm

\centerline{{\it Instituto de F\'\i sica }}
\centerline{\it Universidade Federal do Rio de Janeiro }
\centerline{\it Cx.P. 68528, Rio de Janeiro RJ 21945-970}
\centerline{\it Brazil}

\vskip 0.3cm

\begin{abstract}

Two-dimensional disordered quantum antiferromagnets are studied by means of a continuum
description in which disorder is introduced by a random distribution of couplings
(spin stiffnesses) in the ordered phase of the Nonlinear Sigma Model. 
Quenched soliton (skyrmion)
correlation functions are evaluated and used, along with the quenched magnetization,
to characterize the phase structure of the 
system. When magnetic dilution is exponentially suppressed, 
the introduction of disorder only modifies 
the subleading terms in the large distance behavior of the soliton 
correlation functions, yielding
the same skyrmion energy as in the pure case. The system is in a ``hard'' disordered 
N\'eel phase similar to the ordered antiferromagntic phase occurring in the pure case. 
Conversely, when magnetic dilution is not exponentially suppressed,
the large distance behavior of the correlation functions is
drastically changed. The system exists in a new phase in which the energy of
quantum skyrmions is equal to zero in spite of the existence of a nonvanishing 
antiferromagnetic order parameter. This ``soft'' disordered  N\'eel
phase is characterized by universality classes which are
determined by the behavior of the distribution of random couplings in the small coupling
region. The possible relation of this phase with spin glasses is briefly discussed.

\end{abstract}
\vskip 0.5cm


\vskip 0.5cm
Work supported in part by CNPq, FAPERJ, PRONEX-66.2002/1998-9.

\end{titlepage}

\hoffset= -10mm

\leftmargin 23mm

\topmargin -8mm
\hsize 153mm
 
\baselineskip 7mm
\setcounter{page}{2}

\section{Introduction}
\setcounter{equation}{0}
\bigskip

The continuum description of two-dimensional quantum antiferromagnets has been the object of
intense investigation for a long period of time until quite recently \cite{ha,chn,rs,cn}. 
The interest in this kind of description has been mostly enhanced because of
its successful applications in the
case of layered antiferromagnets such as the high temperature superconducting cuprates. 
For these compounds, the undoped parent materials can be very well described by the
two-dimensional antiferromagnetic Heisenberg model on a square lattice \cite{ak}.
In the continuum limit, this can be mapped into the ordered phase of the 
Nonlinear Sigma Model (NLSM) \cite{ha}, 
whose single coupling constant $\rho_s$, the spin stiffness, is directly related to
the Heisenberg antiferromagnetic coupling $J$. 

Site diluted disordered antiferromagnets have been studied previously in the framework of
the NLSM, leading to very interesting results \cite{cn}.
The aim of this work is to investigate the consequences of disorder in 
the quantum Heisenberg antiferromagnet
by considering a continuous
random distribution of spin stiffnesses
$\rho_s$ in the NLSM.
The effects of disorder are particularly 
interesting and, in fact, lead to unexpected results in the case of skyrmion
correlation functions.
Quantum skyrmion states $| {\rm \ sk}>$ are characterized by the property
$$
Q | {\rm \ sk}> =  | {\rm \ sk}>
$$
where $Q$ is the topological charge operator. 
These excited states are created out of the ground state 
by an operator $\mu$ whose correlation functions and properties have been
extensively studied in \cite{em1}, for the case of the NLSM. 
The skyrmion energy, in particular, can be inferred 
from the large distance behavior of the skyrmion correlation function $<\mu\mu^\dagger>$. 

The pure NLSM at zero temperature is known to exist in two phases \cite{chn}.
These can be characterized by an order parameter, which is the ground state expectation
value of the continuum limit of the sublattice spin operator, namely $<\sigma>$,
along with a dual (disorder) parameter given by $<\mu>$ \cite{em1}. 
One of the phases of the pure system is ordered, having 
$<\sigma> \neq 0$ and $<\mu> = 0$. In this phase $<\mu\mu^\dagger>$ has an 
exponential large distance decay implying the existence of a nonzero 
creation energy for the skyrmions \cite{em1}. 
The other phase is a paramagnetic quantum disordered one, presenting 
$<\sigma> = 0$ and $<\mu> \neq 0$. In this phase, the fact that the skyrmion states
$| {\rm sk}> = \mu |0>$ are not orthogonal to the vacuum mean that there are actually no 
genuine skyrmion excitation in the system. A third possibility, which is not realized in the 
pure NLSM at zero temperature, 
would be a phase in which $<\mu> = 0$ with the skyrmion correlation function 
presenting a power-law decay at large distances. In this case, the system would have
zero energy skyrmions in its excitation spectrum and $<\sigma> = 0$, the absence of order
being closely related to the vanishing of the soliton energy. The possibility of both
$<\mu> \neq 0$ and $<\sigma> \neq 0$, on the other hand, 
is forbidden by a duality relation existing 
between the spin and soliton operators, which has been rigorously demonstrated in one
spatial dimension \cite{km} and, for physical
reasons, should also be valid in higher dimensions.

One can ask whether some different phases may occur when disorder is introduced in the 
system. Starting from the ordered phase of the pure NLSM, which corresponds to
the Heisenberg antiferromagnet, we
investigate this possibility by studying quenched averages in the presence of random
couplings. When magnetic dilution is not exponentially suppressed,
we conclude that the system exists in a new phase, 
where $<<\mu>> = 0 $ and $<<\sigma>> \neq 0$ (the quenched averages $<< >>$ are defined in 
Section 3) but in which the skyrmion
correlation function presents a power-law decay at large distances 
that implies the existence of zero
energy skyrmions.
This phase, which never occurs in the pure system, is a  ``soft'' disordered
N\' eel phase, possessing an order parameter $<<\sigma>> \neq 0$, in spite of the fact that
the skyrmion energy vanishes. The power-law decay of correlation
functions in this phase shows some characteristics of criticality.
Universality classes, which are determined by the behavior of the 
distribution of random couplings at $\rho \rightarrow 0$, also can be clearly identified in
this new phase which bears some resemblance with a spin-glass phase. This is discussed
in Sect.6.

When the distribution 
function of random couplings is such that magnetic dilution is exponentially suppressed, 
on the other hand, 
the system is shown to exist in a phase with $<<\mu>> = 0 $ and $<<\sigma>> \neq 0$ presenting,
at the same time, an exponential decay of the skyrmion 
correlation function, analogously to what happens in the ordered N\' eel phase of the 
pure system. The skyrmion excitations always possess a nonzero energy in this phase. 
Only the subleading behavior of the 
skyrmion and spin correlation functions at large distances is modified by the 
introduction of disorder in this phase, which might be called a ``hard'' disordered
N\' eel phase. 

The paper is organized as follows.
In Section 2, we review some properties of the quantum NLSM relevant for the present work
as well as the continuum limit of two-dimensional quantum antiferromagnets in the
pure case. In Section 3, we consider disorder in the NLSM, manifested in a random distribution
of couplings in the N\'eel phase. The probability distribution functions 
for these couplings are also introduced. In Section 4, 
we study the quenched averages of skyrmion correlation functions in the case where
magnetic dilution is exponentially suppressed. In Section 5, we consider the same 
averages in situations in which magnetic dilution is not suppressed. 
We also show the occurrence of a new phase, presenting some characteristics of
criticality, in which the system belongs to universality classes determined by
the behavior of the distribution function at $\rho_s \rightarrow 0$. Discussion of the results,
conclusions and future perspectives are presented in Section 6.

\section{The Quantum Nonlinear Sigma Model and the Pure Heisenberg Antiferromagnet}

\setcounter{equation}{0}

\subsection{The Quantum Nonlinear Sigma Model}
\bigskip

Let us start by reviewing the properties of the two-dimensional O(3)-symmetric quantum NLSM. 
Subsequently, we shall recall how it is mapped in the 2D Heisenberg antiferromagnet.
The NLSM is defined by the action ($d^3x = d\tau d^2x $)
\be
S = \int d^3x 
\frac{\rho_{0}}{2}
\left[\frac{1}{c^{2}}(\partial_{\tau}{\bf n})^{2} 
+(\nabla{\bf n})^{2}\right],
\label{2}                                             
\ee
where the field ${\bf n}$ is subject to the constraint ${\bf n}^{2}=1$. $\rho_{0}$ 
is a coupling parameter and $c$, a characteristic velocity. Henceforth, unless 
otherwise specified, we shall make $\hbar = 1$ and $c =1$.
Writing the nonlinear sigma field as ${\bf n} = (\sigma, \vec \pi)$, the
zero temperature partition function can be expressed as
$$
Z = \int D \sigma D \vec \pi D \lambda \exp \left \{ - \int d^3 x
\left [ \fr{1}{2}  [(\del_\mu \sigma )^2 + | \del_\mu \vec \pi |^2 ]
\ri . \ri .
$$
\be
\lef . \lef .
+ i \lambda \lef [ \sigma^2 + | \vec \pi | ^2 - \rho_0 \ri ]
\ri ] \ri \},
\label{4}
\ee
where we rescaled the fields and introduced the constraint trough the
Lagrange multiplier field $\lambda$. We use the notation
$\del_\mu \equiv (\fr{\del}{ \del \tau}, \vec \nabla )$.
Integrating on $\vec \pi$, we get the
effective partition function
$$
Z = \int D \sigma  D \lambda \exp \left \{ - \int d^3 x
\left [ \fr{1}{2}  (\del_\mu \sigma )^2 +
i \lambda \lef [ \sigma^2  - \rho_0 \ri ] \ri ]
\ri . 
$$
\be
\lef .
+ tr \ln \lef [ - \Box + i \lambda \ri ] \ri \}
\label{5}
\ee
The constant saddle-point equations derived from the above expression are
$$
<\lambda> <\sigma> = 0
$$
\be
<\sigma>^2 = \rho_0 - \int \fr{d^3k}{(2 \pi)^3} \fr{1}{k^2 + m^2}
\label{6}
\ee
where $\fr{m^2}{2} = i <\lambda> $.
At zero temperature, 
the system presents two phases \cite{chn}: an ordered N\'eel phase, for which
$ <\lambda> = 0$ and $<\sigma>  \neq 0$ and a (paramagnetic) quantum disordered phase,
in which $ <\lambda> \neq 0$ and $<\sigma> = 0$. 
We explore below the physical properties of the
basic excitations of the system and corresponding correlation functions
in each of these two phases.

An important feature of the NLSM is the existence of topological excitations,
called skyrmions. Classically, they are solutions of the field equations 
carrying the topologically conserved charge \cite{bp}
\be
Q = \fr{1}{8\pi} \int d^2 x\epsilon^{ij} \epsilon^{abc} n^a
\del_i n^b \del_j n^c
\label{15}
\ee
At quantum level, the skyrmion states $| {\rm \ sk}> $
are eigenstates of the $Q$ operator with eigenvalue equal to
one and are created by an operator $\mu$ satisfying the commutation rule
$ [Q, \mu ] = \mu $. The correlation functions of this operator have been
studied in detail in the ordered phase of the NLSM, taking into
account full quantum effects \cite{em1}. Together with $<\sigma> $,
the ground state expectation value of the soliton
creation operator $\mu$ is a convenient tool for the characterization of the phases of the 
system, which we are going to exploit.

\subsection{The Quantum Disordered Phase}
\bigskip

We start with 
the quantum disordered phase, where $m \neq 0$.  
Evaluating the integral in (\ref{6})
using the large $k$ cutoff $\Lambda$
and taking $<\sigma> = 0 $, it is easy to see that
\be
\fr{m}{4 \pi} = \fr{\Lambda}{2 \pi^2} - \rho_0 \ > 0
\label{03}
\ee
Note that the large $\Lambda$ behavior of (\ref{03}), as usual, my be compensated by the 
bare coupling $\rho_0$, yielding a finite parameter $m$.
Using the saddle-point solution
$i<\lambda> = \fr{m^2}{2}$, it becomes clear that, up to a constant,
the effective $\sigma$-field action is given by
\be
S_{eff} [\sigma] = \int d^3 x  \fr{1}{2}  \lef [  ( \del_\mu \sigma )^2 +
      m^2 \sigma ^2  \ri ]
\label{8}
\ee
The spin correlation function, in this phase, therefore, is given by
(notice that we are working with imaginary time $\tau = i t$)
$$
<\sigma(\vec x, \tau) \sigma(0, 0)>_{QD} = \int_{-\infty}^{+\infty}
\fr{d\omega}{2\pi} \int  \fr{d^3 k}{(2\pi)^2} 
\fr{e^{i \vec k \cdot \vec x} e^{i\omega \ \tau }
}{\omega^2 + |\vec k|^2 + m^2}
$$
\be
= \fr{e^{- m \sqrt{ |\vec x|^2
+ \tau^2 }}}{4 \pi [  |\vec x|^2 +  \tau^2 ]^{1/2}}
\label{11}
\ee
The exponential decay reveals the presence of a correlation length $\xi = m^{-1}$. The 
large distance behavior $<\sigma \sigma>_{QD} \longrightarrow 0$ confirms 
that $<\sigma>_{QD} = 0$ in this phase. Conversely,
no quantum soliton excitations are expected to present in this phase and therefore we 
must have $<\mu> \neq 0$, implying that the quantum skyrmion state $| {\rm \ sk}> $ 
is not orthogonal to the ground state. As we shall see this will be confirmed below.

\subsection{The Ordered Phase}
\bigskip

We now turn to the ordered phase.
In this case, we have $m = 0$.
Evaluating the integral in (\ref{6}), again using the large $k$ cutoff $\Lambda$, we get
\be 
<\sigma>_{ORD}^2 = \lef [ \rho_0 - \fr{\Lambda}{2 \pi^2} \ri ]\  > 0
\label{01}
\ee
Once more, the large $\Lambda$ behavior in (\ref{01})
can absorbed in a redefinition of the bare coupling $\rho_0$. 
Introducing the renormalized (finite) coupling $\rho_s$, the spin stiffness, as
\be
\rho_s = \rho_0 - \fr{\Lambda}{2 \pi^2}\ \geq 0
\label{02}
\ee
we see that $<\sigma>_{ORD}^2 = \rho_s$, which is nonzero in the ordered phase. 
At the quantum critical point
$\rho_s = 0$, the system enters the disordered phase, where $m \neq 0$. The sublattice
magnetization $M$ is given by $M = <\sigma> $.
In the ordered phase, we have $M_{ORD} = \sqrt{\rho_s}$ while
in the quantum disordered phase, of course, $M_{QD} = 0$.

Let us consider now the complete renormalization of the theory in the ordered phase. 
From (\ref{02}), we can write
\be
\rho_0 = Z \rho_s \ \ \ ;\ \ \ Z = \lef (1 + \fr{\Lambda}{2 \pi^2 \rho_s}\ri )
\label{04}
\ee
Introducing the renormalized fields ${\bf n}_R, \lambda_R$ and action $S_R$ through
\be
{\bf n} = Z^{-1/2}\ {\bf n}_R \ \ ; \ \ \lambda =  Z\ \lambda_R\ \ ;\ \ S_R = S + \delta S
\label{05}
\ee
where
\be
\delta S = - i (Z^{-1} - 1) \int d^3x \ Z \lambda_R 
\label{06}
\ee
it is easy to see that the renormalized action is given by
\be
S_R = \int d^3x \lef \{
\frac{\rho_{s}}{2}
| \partial_{\mu}{\bf n}_R |^{2} + i \lambda_R \lef [ |{\bf n}_R |^{2} - 1  \ri ]  \ri \}
\label{07}                                             
\ee
This is identical to the classical action, but with renormalized, physical quantities, 
replacing the bare ones. For this reason, 
this phase in known as ``renormalized classical'' \cite{chn}. We see, 
in particular, that the physical coupling constant in this phase is the 
spin stiffness $\rho_s$.

Replacing $S$ for $S_R$ in (\ref{5}), inserting the saddle-point value $<\lambda_R> =0$ 
and shifting the $\sigma_R$-field around its vacuum expectation value
$<\sigma_R> = \sqrt{\rho_s}$, namely defining
$  \eta \equiv \sigma_R - \sqrt{\rho_s} $ we get
\be
S_{eff} [\eta] = \int d^3 x  \fr{1}{2} ( \del_\mu \eta )^2  
\label{13}
\ee
This is the well-known Goldstone boson action and the corresponding correlation functions are
\be
<\eta(\vec x, \tau) \eta(0, 0)> =
<\sigma_R(\vec x, \tau) \sigma_R(0, 0)>_{OAF} - <\sigma_R>_{OAF}^2 =
\fr{1}{4 \pi [  |\vec x|^2 +  \tau^2 ]^{1/2}}.
\label{14}
\ee
We now see that $<\sigma_R \sigma_R>_{OAF} \longrightarrow <\sigma_R>_{OAF}^2 \neq 0 $,
at large distances,
thus confirming the fact that  $ <\sigma_R>_{OAF} \neq 0 $ in this phase.

In the ordered antiferromagnetic phase, we have the occurrence of 
classical skyrmion excitations, possessing $Q=1$. These are given by \cite{bp}
\be
\vec n_S (\vec x) = \rho_s \lef ( \sin f(r) \hat r, \cos f(r) \ri )
\label{15}
\ee
with 
$$
f(r) = 2 \arctan \fr{l}{r}
$$
where $l$ is an arbitrary scale and $r$ is the radial distance in
two-dimensional space.
The energy of this classical skyrmion excitation in the ordered phase, described by the
renormalized classical action (\ref{07}) and measured
with respect to the ordered antiferromagnetic background 
is $ E = 4 \pi \rho_s $. This must compared
with the full quantum result obtained from a quantized skyrmion field theory.
The two-point quantum skyrmion correlation
function has been evaluated in the ordered phase of the quantum NLSM \cite{em1} and 
the result is
\be
<\mu(\vec x , \tau) \mu^{\dagger}(0, 0)>_{OAF} =
\exp \lef \{- 2 \pi \rho_s [   |\vec x|^2 + \tau^2 ]^{1/2} \ri \}
\label{16}
\ee
From this we can infer that the actual energy of the full quantum
skyrmions in the ordered antiferromagnetic phase is $ 2 \pi \rho_s $, that
is, a half of the classical value. From (\ref{16}), we also see that
$<\mu \mu^\dagger>_{OAF} \longrightarrow  0 $, at large distances, implying
that $<\mu> = 0 $ in this phase. 
This means that the quantum skyrmion state $| {\rm \ sk}> $ is orthogonal to the vacuum
and the quantum skyrmions are genuine excitations. For $\rho_s \rightarrow 0$, on the
other hand, when we approach the quantum critical point leading to the disordered phase,
we see from (\ref{16}) that $<\mu> \neq 0 $, confirming therefore our anticipation for
the ground state expectation value of the skyrmion operator in the disordered phase.

\subsection{Connection with the Heisenberg Antiferromagnet}
\bigskip

Two-dimensional antiferromagnets on a square lattice can be described by the
O(3)-symmetric Heisenberg hamiltonian, given by
\be
H =  \sum_{<ij>} J_{ij} \vec S_i \cdot \vec S_j
\label{1}
\ee
where the sum runs only over nearest neighbor sites and $ J_{ij}  > 0 $.
The ``pure'' case is characterized by the fact that
the coupling constants $ J_{ij} $ are determined and fixed.
In the homogeneous case, all the coupling constants are equal
and we have $ J_{ij} \equiv J > 0 $. 
At zero temperature, this system is known to exist only in an ordered N\'eel phase. The
quantum fluctuations are not capable of destroying the long-range antiferromagnetic order
for any value of the coupling constant \cite{ma}.

It has been shown that 
in the continuum limit, the above quantum hamiltonian, in the homogeneous case,
is mapped into the {\it ordered phase} of the
quantum NLSM \cite{ha,chn}, the nonlinear sigma field ${\bf n}(\vec x, t)$ being the continuum
limit of the sublattice spin operator. The spin stiffness $\rho_s$ which, as we saw, 
controls the whole physical properties of the system 
in the ordered phase, is related to the Heisenberg
antiferromagnetic coupling $J$ as \cite{chn,ak}
\be
\rho_s = J S^2 Z_{\rho_s}
\label{08}
\ee
where $S$ is the spin quantum number and $Z_{\rho_s}$ is a constant accounting for 
quantum corrections to the classical continuum limit. For $ S = \fr{1}{2}$, we have
\cite{chn,ak}
\be
\rho_s \simeq 0.18 \ J
\label{09}
\ee

In the non-homogeneous case, where the coupling constants $J_{ij}$ are different for
each link, we
can derive the continuum limit by following the same procedure as in \cite{ha},
provided the configuration of coupling constants $J_{ij}$ in (\ref{1}) is
slowly varying (this is going to be made precise in what follows). In this case, we obtain
in the continuum limit,
a Nonlinear Sigma Model with the spin stiffness
$\rho_s$ replaced by a slowly varying configuration $\rho(\vec r)$ that is
related to $ J_{ij} $ in the same way that $\rho_s$ is related to $J$, namely
\be
S = \int d^3x \lef \{
\frac{1}{2}
| \partial_{\mu}{\bf n} |^{2} + i \lambda \lef [ |{\bf n} |^{2} - \rho_0(\vec r)\ri ] \ri \}
\label{010}                                             
\ee
This is equivalent to modifying the constraint from $\delta[ |{\bf n} |^{2} - 1]$
to $\delta[ |{\bf n} |^{2} - f(\vec r)]$, with $\rho_0(\vec r) \equiv \rho_0 f(\vec r)$. 
Integrating over $\vec \pi$, we get
\be
S = \int d^3x \lef \{
\frac{1}{2}
( \partial_{\mu} \sigma)^{2} + i \lambda [ \sigma^{2} - \rho_0(\vec r)] \ri \}
+  tr \ln \lef [ - \Box + i \lambda \ri ] 
\label{011}                                             
\ee
Now translation invarianace is lost and the saddle-point equations become
$$
- \nabla^2 <\sigma>  + 2 m^2(\vec r) <\sigma> = 0
$$
\be
<\sigma>^2 = \rho_0(\vec r) - \int \fr{d^3k}{(2 \pi)^3} \fr{1}{k^2 + m^2(\vec r)} \equiv
\rho(\vec r)
\label{012}
\ee
We see that now $<\sigma> = <\sigma>(\vec r)$ and consequently $m(\vec r) \neq 0$, in spite of 
the fact that $<\sigma> \neq 0$. The spin correlation function becomes damped now, with a
damping factor $m(\vec r)$. This is in agreement with previous investigations of 
spin-waves in similar situations \cite{cn,sw}. 

The soliton correlation function can be evaluated in a saddle-point approximation 
in the theory described by (\ref{010}) \cite{em1}. This has a simple expression in terms
of \\ $<\sigma>$, which will be very convenient for the obtainment of quenched averages in
the disordered version of the model, namely,
\be
<\mu(\vec x , \tau) \mu^{\dagger}(0, 0)> =
\exp \lef \{- 2 \pi <\sigma>^2 [ |\vec x|^2 + \tau^2 ]^{1/2} \ri \} =
\exp \lef \{- 2 \pi \rho(\vec r) [ |\vec x|^2 + \tau^2 ]^{1/2} \ri \}
\label{013}
\ee

In the rest of this work, we consider the situation in which
intrinsic disorder is introduced in the system and investigate its effects on the soliton 
correlation function.

\section{The Continuum Limit of Disordered Antiferromagnets}

\setcounter{equation}{0}

\subsection{The Disordered System}

\bigskip

Let us describe the presence of disorder in the two-dimensional 
Heisenberg quantum antiferromagnet, 
by considering
a random distribution of couplings $ J_{ij} $ in (\ref{1}), in analogous 
way as in the
Edwards-Anderson model \cite{ea}. Here, however, we will keep only
antiferromagnetic couplings $ J_{ij} > 0 $. This will allow us to easily
obtain a continuum field theory version for the disordered model,
in the same way as in the pure case.
The disorder is introduced in the continuum version by taking a
random distribution $ P[\rho (\vec r) ]$ for the
slowly varying spin stiffness $\rho (\vec r) $ appearing in (\ref{010}).
We require that
$ P [\rho (\vec r) ] = 0$ for
$\rho (\vec r) < 0 $. This will ensure that in spite of the presence of
disorder, we are always in the ordered phase of the NLSM, for which the mapping to
the Heisenberg antiferromagnet exists.
We also impose the condition that the variance of
this distribution is always much smaller than $\rho_s$, in order to
ensure that the $\rho (\vec r)$-configurations are slowly varying.
    
We are only going to consider the quenched case and
take quantum averages at zero temperature using a fixed
configuration for $\rho (\vec r) $.
Subsequently we shall evaluate the average over the
$\rho (\vec r)$-configurations using the $ P[\rho (\vec r) ]$
distribution function.
The relevant average for an operator $\ca$ in the quenched random system
will be therefore
\be
<<\ca>_q>_\rho = \fr{1}{\Lambda} \Pi_{\vec r}
\int_0^\infty d\rho (\vec r)
P[\rho (\vec r)] <\ca>_q (\rho (\vec r))
\label{17}
\ee
where $\Lambda \equiv (\Pi_{\vec r} \cdot 1)$
is a normalization factor corresponding,
in the lattice, to a product over all links $(ij)$ and $<\ca>_q$ is the zero temperature 
quantum average.

\subsection{Distribution Functions}
\bigskip

Let us introduce now the distribution functions we are going to use,
in order to describe the disorder. We shall consider basically two
functions containing a gaussian distribution centered around the pure
spin stiffness $\rho_s $. The first one is (henceforth we omit the argument $\vec r$
in $\rho$)
\be
P_1[\rho] =
 \lef \{  \begin{array}{c}
\fr{1}{N_1}  (\rho - \rho_s)^{\nu -1}
 e^{- \fr{(\rho - \rho_s)^2}{2 \Delta^2}}\ \
 \rho > \rho_s  \\    \\

 \fr{1}{N_1} (\rho_s - \rho)^{\nu -1}
 e^{- \fr{(\rho - \rho_s)^2}{2 \delta^2}}\ \
 0 \leq \rho < \rho_s  \\    \\

 0 \ \ \ \ \ \ \ \ \ \ \ \ \ \ \ \ \ \ \ \ \ \ \  \rho < 0
          \end{array} \ri .
\label{18}
\ee
where $\nu > 0$.
In this expression, we assume that both $\Delta << \rho_s$ and
$\delta << \rho_s$, thereby guaranteeing that the random
$\rho$-configurations are slowly varying. We also assume that
\be
 \lef ( \fr{L}{\hbar c}\ri ) \Delta >> 1
\label{19}
\ee
where $L$ is the maximum dimension of the system and $c$, the
spin-wave velocity, its characteristic velocity. Two regimes of disorder
described by (\ref{18}) can be distinguished and will produce a completely
different behavior of the correlation functions, as we shall see.
A first one is obtained by choosing a symmetric gaussian, with
$\delta = \Delta$. A second one with the choice $\delta << \Delta$ and
\be
 \lef ( \fr{L}{\hbar c}\ri )  \delta << 1
\label{20}
\ee
In the second case, there is a severe exponential suppression of values
of the spin stiffness around $\rho_s = 0$, that is, magnetic dilution is
exponentially suppressed in a very strong way. In the first case, dilution
is not so much suppressed and, as we shall see, the system has the same
qualitative behavior as a diluted one.

Experimental values for the parameters of the above distribution,
in the case of high-temperature superconducting cuprates in the ordered
antiferromagnetic phase, which are typical examples of two-dimensional
Heisenberg antiferromagnets are \cite{ak}:
$\rho_s \simeq 10^{-1} eV$, $\hbar c \simeq 1 eV \AA $. For a sample of
dimension $L \simeq 1mm$, we choose
$\Delta \simeq 10^{-3}eV$, which satisfies
(\ref{19}). For $\delta$, we have
$\delta = \Delta$ in the case with dilution. In the case without dilution
the choice $\delta \simeq 10^{-9}eV$ will satisfy (\ref{20}). In both cases,
the condition $\Delta, \delta << \rho_s$ is satisfied.

In the distribution function (\ref{18}), the normalization factor is given by
\be
N_1 = 2^{\fr{\nu}{2} -1} \Gamma\lef (\fr{\nu}{2}\ri )
\lef [\Delta^{\nu} + \delta^{\nu}  \ri ]
\label{21}
\ee
The average spin stiffness is
\be
\bar \rho = \rho_s + 
\fr{\Gamma \lef ( \fr{\nu +1}{2} \ri )}{\sqrt{2}\Gamma
\lef ( \fr{\nu}{2} \ri )} \ \ \Delta 
\label{22}
\ee
in the case where dilution is not suppressed and
\be
\bar \rho = \rho_s + \sqrt{2}
\fr{\Gamma \lef ( \fr{\nu +1}{2} \ri )}{\Gamma
\lef ( \fr{\nu}{2} \ri )} \ \ \Delta
\lef [ 1 - \lef(\fr{\delta}{\Delta}  \ri )^{\nu}  \ri ]
\label{23}
\ee
in the presence of exponential suppression of dilution ($\delta << \Delta$).

The second distribution function we are going to use is
\be
P_2[\rho] =
 \lef \{  \begin{array}{c}
\fr{1}{N_2}  \rho^{\nu -1}
 e^{- \fr{(\rho - \rho_s)^2}{2 \sigma^2}}\ \
 \rho \geq 0  \\    \\
 0 \ \ \ \ \ \ \ \ \ \ \ \ \ \ \ \ \ \ \ \ \ \ \  \rho < 0
          \end{array} \ri .
\label{23}
\ee
where $\nu > 0$. We assume $\sigma << \rho_s$, again to ensure that the
$\rho$-configurations are slowly varying.
This condition can be experimentally
satisfied, in the case of the high-temperature cuprates, with a choice of
$\sigma \simeq 10^{-3}eV$. Observe also that using the same
experimental values of the previous paragraph, this value of $\sigma$
satisfies the condition
\be
 \lef ( \fr{L}{\hbar c}\ri ) \sigma >> \lef (\fr{\rho_s}{\sigma} \ri ) >> 1
\label{24}
\ee
which is similar to (\ref{19}).

The normalization factor in (\ref{24}) is now given by
\be
N_2 = \sigma^\nu \Gamma(\nu) D_{-\nu}(-\fr{\rho_s}{\sigma})
e^{-\fr{\rho_s^2}{4\sigma^2}}
\label{25}
\ee
where $D_{-\nu}(x)$ is a parabolic cylinder function.
For the $P_2[\rho]$ distribution function, the average spin stiffness is
\be
\bar \rho = \rho_s + (\nu -1) \fr{\sigma^2}{\rho_s}
\label{26}
\ee

\section{Disorder With Exponentially Suppressed Dilution}

\setcounter{equation}{0}

\bigskip

In this section, we consider the situation in which 
magnetic dilution is exponentially suppressed in the disordered system. As explained above,
this correponds to the choice of $P_1[\rho]$ as the distribution function , with the parameters
$\Delta$ and $\delta$ satisfying (\ref{19}) and (\ref{20}), respectively. We are going to 
evaluate the quenched skyrmion correlation functions,
starting from the ordered antiferromagnetic 
phase of the pure system. In this case, the zero temperature 
pure quantum averages 
$<\mu \mu^{\dagger}>_{OAF}$, given by (\ref{16}), depend
exponentially on the spin stiffness. For the disordered NLSM, introduced in subsection 2.2,
in the regime where the spin stiffness $\rho (\vec r)$ is slowly varying, the soliton 
correlation fubction is given by (\ref{013}). 
Hence, when evaluating the quenched averages (\ref{17}),
we shall have a zero temperature 
quantum average whose $\rho (\vec r)$-dependence is of the form
$\exp \{ - \alpha \rho (\vec r) \}$
where $\alpha = 2 \pi X $, with 
$X \equiv [ |\vec x|^2 +  \tau^2 ]^{1/2}$.
The relevent integral for the evaluation of (\ref{17}) is, therefore
\be
A = A_1 + A_2
\label{28}
\ee
where
\be
A_1 = \fr{1}{N_1} \int_0^{\rho_s} d \rho (\rho_s - \rho )^{\nu - 1} e^{-\alpha \rho}
e^{- \fr{(\rho - \rho_s)^2}{2 \delta^2}} \\
= \fr{1}{N_1 \alpha^\nu} \int_0^{\alpha\rho_s} d x (\alpha\rho_s - x )^{\nu - 1} e^{- x}
e^{- \fr{(x - \alpha\rho_s)^2}{2 \alpha^2\delta^2}} 
\label{29}
\ee
and
\be
A_2 = \fr{1}{N_1} \int_{\rho_s}^\infty d \rho (\rho - \rho_s )^{\nu - 1} 
e^{-\alpha \rho}
e^{- \fr{(\rho - \rho_s)^2}{2 \Delta^2}} \\
= \fr{e^{-\alpha \rho_s}}{N_1 \alpha^\nu} \int_0^{\infty} d x x^{\nu - 1} 
e^{- x}
e^{- \fr{x^2}{2 \alpha^2\Delta^2}} 
\label{30}
\ee
We shall be interested in the behavior of quenched averages
of correlation functions, given by (\ref{17}),
at large distances ($X \rightarrow \infty$). In this case we have 
$\alpha \rightarrow \infty$ and we can, therefore, use conditions
(\ref{20}) and (\ref{19}), respectively, in (\ref{30}) and (\ref{29}), 
to obtain
\be
A_1 \stackrel{\alpha \rightarrow \infty}{\longrightarrow}
\fr{e^{-\alpha \rho_s}}{N_1 \alpha^\nu}
\int_{\alpha\rho_s - \alpha\delta}^{\alpha\rho_s} d x 
(\alpha\rho_s - x )^{\nu - 1}
= \fr{\delta^\nu}{N_1 \nu } e^{-\alpha \rho_s}
\label{32}
\ee
and
\be
A_2 \stackrel{\alpha \rightarrow \infty}{\longrightarrow}
\fr{e^{-\alpha \rho_s}}{ N_1\alpha^\nu}
\int_0^{\infty} d x x^{\nu - 1} e^{- x}
= \fr{\Gamma (\nu)}{N_1 } \fr{e^{-\alpha \rho_s}}{\alpha^\nu}
\label{31}
\ee
(we could also have obtained the result in (\ref{31}) by exactly evaluating the last integral 
in (\ref{30})  
and subsequently taking the limit $\alpha \rightarrow 0$ (see (\ref{43}) and (\ref{44}))).
We conclude from (\ref{32}) and (\ref{31}) that
\be
A \stackrel{\alpha \rightarrow \infty}{\longrightarrow}
\fr{e^{-\alpha {\rho_s}}}{N_1} \lef [ \fr{\Gamma (\nu)}{\alpha^\nu } +
\fr{\delta^\nu}{\nu }  \ri ] \\
A \stackrel{\alpha \rightarrow \infty}{\longrightarrow}
\fr{\Gamma (\nu)}{N_1 } \fr{e^{-\alpha {\rho_s}}}{\alpha^\nu}
\label{33}
\ee
where we used (\ref{20})in the second line, 
in order to get the dominant behavior for large $\alpha$.

From (\ref{16}) and (\ref{33}) we can immediately obtain the
expression for the long distance behavior of the skyrmion
quenched averages in the absence of magnetic dilution, namely, 
\be
<<\mu \mu^{\dagger}>>_{OAF}\stackrel{X \rightarrow \infty}{\longrightarrow}
\fr{\Gamma (\nu)}{(2 \pi)^\nu N_1}
\lef ( \fr{e^{- 2 \pi {\rho_s} X}}{X^{\nu}}\ri ).
\label{35}
\ee
Observe that the presence of disorder, in the case where dilution is esponentially
suppressed, does not
modify the dominant exponential large distance behavior. Only the 
subdominant power law decay is modified by the introduction of an
additional exponent $\nu$.

\section{Disorder Without Exponential Suppression of Dilution}

\setcounter{equation}{0}

\bigskip

In this section, we are going to consider the situation in which 
magnetic dilution is not exponentially suppressed by the random distribution of
couplings. As we saw in Sec. 2, this can happen either when we use
the distribution $P_1 [\rho]$ in the symmetric case 
when $\delta = \Delta $ or when we use $P_2 [\rho]$. In what follows, we study the two cases
separately.

\subsection{Distribution Function $P_1 [\rho]$}
\bigskip

In this case, the relevent integral for the evaluation of (\ref{17}) is
\be
B = B_1 + A_2
\label{36}
\ee
where $A_2$ is given by (\ref{30}) and 
\be
B_1 = \fr{1}{N_1} \int_0^{\rho_s} d \rho (\rho_s - \rho )^{\nu - 1} e^{-\alpha \rho}
e^{- \fr{(\rho - \rho_s)^2}{2 \Delta^2}} \\
=  \fr{e^{-\alpha \rho_s}}{N_1 \alpha^\nu} \int_0^{\alpha\rho_s} d x
x^{\nu - 1} e^{x} e^{- \fr{x^2}{2 \alpha^2\Delta^2}} 
\label{37}
\ee
In the large distance regime, when $\alpha \rightarrow \infty$, we can
use (\ref{19}) to obtain \cite{gr}
\be
B_1 \stackrel{\alpha \rightarrow \infty}{\longrightarrow}
\fr{e^{-\alpha {\rho_s}}}{N_1 \alpha^\nu} \int_0^{\alpha{\rho_s}} d x
x^{\nu - 1} e^{x} = \fr{\rho_s^\nu\ e^{-\alpha {\rho_s}}}{\nu \ N_1}\ \ 
_1F_1 (\nu; 1+\nu; \alpha \rho_s)
\label{38}
\ee
where $ _1F_1 (\nu; 1+\nu; \alpha \rho_s)$ is a confluent hypergeometric 
function. Using the large distance asymptotic behavior of this function,
we get \cite{gr}
\be
B_1 \stackrel{\alpha \rightarrow \infty}{\longrightarrow}
\fr{\rho_0^{\nu - 1}}{N_1 \alpha}
\label{39}
\ee
Combining (\ref{39}) with (\ref{33}), we see that
\be
B \stackrel{\alpha \rightarrow \infty}{\longrightarrow} \fr{1}{N_1}
\lef [ \fr{\rho_s^{\nu - 1}}{\alpha} +
\Gamma (\nu)\fr{e^{-\alpha \rho_s}}{\alpha^\nu} \ri ] 
 \stackrel{\alpha \rightarrow \infty}{\longrightarrow}
\fr{\rho_s^{\nu - 1}}{N_1 \alpha}
\label{40}
\ee
From this result, we can immediately infer the large distance 
behavior of the skyrmion quenched correlation function. This is
given by
\be
<<\mu \mu^{\dagger}>>_{OAF}\stackrel{X \rightarrow \infty}{\longrightarrow}
\fr{\rho_s^{\nu -1}}{ 2 \pi N_1 }
\lef ( \fr{1}{X} \ri ).
\label{42}
\ee
Now the large distance behavior of the correlation function is 
drastically changed. The previously dominating exponential decay is 
completely washed out and we have, instead, a power-law decay. As
we shall argue in the next subsection, the exponent of the power-law is universally
determined by the behavior of the distribution function at 
$\rho \rightarrow 0 $.

\subsection{Distribution Function $P_2 [\rho]$}

\bigskip

Let us consider now the situation in which the disorder is described
by the distribution function $P_2 [\rho]$. The relevant integral for
the evaluation of (\ref{17}) is now \cite{gr}
$$
C = \fr{1}{N_2} \int_0^\infty d \rho  \rho^{\nu - 1} e^{-\alpha \rho}
e^{- \fr{(\rho - \rho_s)^2}{2 \sigma^2}} \\
= \lef [D_{-\nu} \lef (-\fr{\rho_s}{\sigma}\ri )\ri ]^{-1}
\exp \lef [ - \fr{ \rho_s \alpha}{2} \ri ] 
$$
\be
\times\  
\exp \lef [ \fr{ \sigma^2 \alpha^2 }{4}\ri ]
\ \  D_{-\nu}\lef( \sigma  \alpha - \fr{\rho_s}{\sigma}  \ri )
\label{43}
\ee
where $D_{-\nu}(x)$ is the parabolic cylinder function. The large 
distance behavior of (\ref{43}) can be obtained by considering
the property \cite{gr}
$$
D_{-\nu}(x) \stackrel{x \rightarrow \infty}{\longrightarrow}
e^{- \fr{x^2}{4}} x^{-\nu}
$$
\be
D_{-\nu}(-x) \stackrel{x \rightarrow \infty}{\longrightarrow}
\fr{\sqrt{2\pi}}{ \Gamma(\nu)} e^{ \fr{x^2}{4}} x^{\nu-1}
\label{44}
\ee
Using this and (\ref{24}), we can immediately obtain the large
distance behavior of the quenched skyrmion correlation 
functions for the distribution function $P_2 [\rho]$. This is
given by
\be 
<<\mu \mu^\dagger>>
\stackrel{X \rightarrow \infty}{\longrightarrow}
\fr{\Gamma(\nu)}{ (2\pi)^{\nu + 1/2} }
\lef(\fr{\rho_s^{1-\nu}}{\sigma} \ri )
\lef ( \fr {1}{ X^{\nu}} \ri )
\label{46}
\ee
We observe here that, as in the case of the distribution used in the previous subsection, the
introduction of disorder completely modifies the large distance
behavior of the correlation functions, eliminating the exponential decay.
This fact can be generally understood by observing that the large
distance (large $\alpha$) behavior of (\ref{17})
is determined by the behavior of the distribuiton $P [\rho]$ at 
$\rho \rightarrow 0$. This happens because, for the distribution functions
$P [\rho]$ (with support in the region $\rho \geq 0$) and quantum averages $<\ca>_q (\rho )$
(exponentially depending on $\rho$) considered in this work, the quenched average (\ref{17}) 
is proportional to the $\alpha$ Laplace transform of $P [\rho]$.
As a consequence, the asymptotic large distance behavior of (\ref{17}) is universally 
determined by the behavior of $P [\rho]$ for
$\rho \rightarrow 0$. We see, for instance, that the values of the parameter $\nu$ in 
$P_2 [\rho]$ determine universality classes to which the disordered system belongs. 
This can be confirmed by the behavior of the correlator (\ref{42}).
Also, by comparing (\ref{42}) with (\ref{46}) 
we conclude that distribution $P_1 [\rho]$, in the symmetric case when $\delta = \Delta $, 
is in the  $\nu = 1$ universality class of 
$P_2 [\rho]$. This can be also verified by observing that both 
distributions have the same type of behavior at $\rho \rightarrow 0$. Conversely, in the 
asymmetric case when $\delta << \Delta $, the distribution function 
$P_1 [\rho]$ is exponentially suppressed at $\rho \rightarrow 0$, a type
of behavior which is never presented by $P_2 [\rho]$, leading to an exponential decay of the
quenched correlators.

\section{Discussion and Conclusions}

\setcounter{equation}{0}

\bigskip

Our continuum analysis of two-dimensional quantum antiferromagnets in the presence of a random 
distribution of couplings at zero temperature,
has shown that the quenched averages of soliton (skyrmion) 
correlation functions are modified with respect to the pure case. The modification
is particularly drastic whenever magnetic dilution is not exponentially suppressed 
in the disordered system.
This effect has quite interesting consequences in the physical properties of skyrmions.
It is a well known fact that an exponential decay of the
soliton correlation function at large distances
would indicate that the energy of the soliton excitations is nonzero and
proportional to the coefficient of the exponent. In the pure NLSM in the ordered phase, this is
given by $E_S = 2 \pi \rho_s = 2 \pi <\sigma>^2 $. A nonzero soliton energy, therefore,
is associated to an ordered ground state with $<\sigma> \neq 0$. Physically this can
be understood as a consequence of the fact that 
in an ordered ground state there is an energy cost to make the
spin flips necessary for the introduction of a soliton state. The exponential decay of
$<\mu\mu^\dagger>$, further implies through 
$<\mu\mu^\dagger> \rightarrow |<\mu>|^2 $ that $<\mu> = 0$ which means that the soliton
states are orthogonal to the vacuum, that is to say, true excitations. A power-law decay,
on the other hand, while still leading to $<\mu> = 0$ and therefore
meaning that quantum solitons are genuine excitations 
would imply that the energy necessary for the creation of these solitons is equal to zero. 
In a {\it generic pure} system at zero temperature, 
this would correspond to a {\it quantum} disordered ground state
($<\sigma>=0$), because when the ground state is not an ordered one, 
there is no energy cost for introducing the spin flips necessary to create a soliton state. 
These zero energy quantum skyrmions, in spite of bearing few relation with their classical
ancestors occurring in an ordered phase, would
exist as true physical excitations in such a phase.
It should be stressed, however, that the kind of quantum disorder occuring when we have a 
power-law decay of the soliton correlation function, differs from the one found 
in a paramagnetic phase, such as the quantum disordered phase of the NLSM, in which
we have $<\mu\mu^\dagger> \rightarrow C \neq 0$, that is
$<\mu> \neq 0$ and $<\sigma> = 0$. Consequently, we 
conclude that a power-law decay of the soliton
correlation function would imply some different type of disorder than the one found in
a paramagnetic phase. 
In summary we have the following possibilities for the phases 
of a pure quantum antiferromagnet:
\be
         \begin{array}{c}
  <\mu\mu^\dagger>
  \stackrel{X \rightarrow \infty}{\longrightarrow}
  C \neq 0 \ \ \ ; \ \ \   <\mu> \neq 0, \ <\sigma> = 0\ \
  {\rm  - \ \  Paramagnetic\ \ Quantum\  Disordered}
  \\   \\ 
  <\mu\mu^\dagger>
  \stackrel{X \rightarrow \infty}{\longrightarrow} 
  e^{-E_S X} \ \ \ ; \ \ \  <\mu> = 0, \ <\sigma> \neq 0\ \
  {\rm  - \ \ Antiferromagnetic\ (N\acute eel)}
  \\   \\ 
  <\mu\mu^\dagger>
  \stackrel{X \rightarrow \infty}{\longrightarrow}
  \fr{1}{X^\nu}  \ ; \   <\mu> = 0, \
  <\sigma> = 0\ \
  {\rm  - \ \  Non\ Paramagnetic\ Quantum\  Disordered}
          \end{array} 
\label{47}
\ee
The first two ones are realized in the pure NLSM \cite{chn} at zero temperature and only the 
second one in the pure two-dimensional Heisenberg quantum antiferromagnet with
nearest neighbors interaction, also at $T =0$ \cite{ma}.

We then consider the presence of disorder.
Starting from the ordered antiferromagnetic phase of the pure NLSM, which corresponds
to the Heisenberg antiferromagnet,
we have studied in this work the 
effects of disorder introduced through a continuum random distribution of couplings. 
In all cases considered here, the quenched magnetization is nonzero, namely,
$M_Q = <<\sigma>> \simeq \sqrt{\bar \rho}$, where $\bar \rho$ is given by (\ref{22})
(\ref{23}) and (\ref{26}), respectively.
The studied systems are always in an antiferromagnetic 
``disordered'' N\' eel phase having 
a nonzero order parameter in spite of the fact that the couplings are random.
The types of disorder considered here, 
{\it ab initio} cannot destroy the antiferromagnetic order of the pure system 
since they only allow the presence of positive or null couplings. 
There are,
however, two possible types of such phases, which we call ``soft'' or ``hard'',
according to whether magnetic dilution is exponentially suppressed or not.

When dilution is exponentialy suppressed, 
only the subleading term of the correlation functions at
large distances is modified by disorder. The exponential decay of the pure system is 
preserved and the skyrmion energy in the disordered system is the same as in the pure 
case. The system exists in a phase corresponding to the second possibility in (\ref{47}),
which may be called a ``hard'' disordered N\'eel phase.
Conversely, when dilution is not exponentially suppressed, the large distance behavior
is drastically changed from an exponential to a power-law decay. The skyrmion energy
consequently becomes zero in the disordered system, despite the fact that the magnetization 
is non-vanishing and $<<\sigma>> \neq 0$. This is a completely novel behavior for quantum 
skyrmions. Physically, we can understand it as follows: the disorder introduced by the 
random distribution of couplings is sufficient to reduce to zero the energy necessary for 
the creation of a skyrmion configuration, even though it cannot destroy the 
quenched magnetization. The soliton correlation functions have a universal
behavior, characterized by universality classes which are determined by the 
behavior of the distribution function at $\rho_s \rightarrow 0 $.

It would be interesting to find a physical realization for this phase in which 
skyrmion correlators have a 
power-law decay at large distances and which is
associated to a diluted disordered antiferromagnet. One might be tempted to associate it 
to a spin-glass phase since spin-glasses can be obtained by dilution of
antiferromagnetic systems. A well known example
in three dimensions is the substance $ZnCr_{2-x}Ga_xO_4$,
obtained by diluting the pure antiferromagnet $ZnCr_2O_4$ with the nonmagnetic
$Ga$ atoms \cite{daf,sg1}. The possibility of the ``soft'' phase, obtained in the present
work by dilution of a pure antiferromagnet, being a spin-glass,
however, is ruled out by the fact that the quenched magnetization is nonvanishing.
A key ingredient for this would be the presence of 
frustration which is absent in the system 
studied here but, conversely, present in the case of $ZnCr_{2-x}Ga_xO_4$.
The diluted phase studied here is actually a new one which could be added to the list
(\ref{47}) and is characterized by
\be
 \begin{array}{c}
  <\mu\mu^\dagger>
  \stackrel{ X \rightarrow \infty}{\longrightarrow}
  \fr{1}{X^\nu}  \ \ \ ; \ \ \   <\mu> = 0, \ <\sigma> \neq 0\ \
 {\rm  - \ \   Soft\ Disordered\ N\acute eel}
   \end{array} 
\label{49}
\ee
In this phase, the skyrmion energy vanishes, not as a consequence of quantum fluctuations,
as in the third phase listed in (\ref{47}), but rather, because of the type of disorder
introduced when magnetic dilution is not exponentially suppressed. This, in spite of not being
capable of destroying the order parameter, forces the skyrmion energy to vanish.
The physical properties of the zero energy quantum solitons 
occuring in these kind of phases may play an important role in planar 
antiferromagnetic systems such 
as high temperature superconducting cuprates an shall be exploited elsewhere.
For this purpose, one
should investigate the effects of disorder in the continuum models
for doped antiferromagnetic planar systems used to describe materials 
such as LSCO and YBCO \cite{em2}.

Let us stress, finally, that that the kind of magnetic dilution considered in this work,
even in the cases where it is not exponentially suppressed, 
is always softer than a delta function-type of dilution containing a
piece $P[\rho] = x \delta (\rho)$, which would be suitable for describing the type
of dilution occuring in compounds such as $La_2 Cu_{1-x}Zn_x O_4$ \cite{cn}.
As future extensions of this work, we  
intend to consider more general disorder distributions
including these and ferromagnetic couplings as well. 
In the latter case, we can expect to describe spin-glass 
phases as well. It is likely that in a spin-glass phase 
the behavior of soliton correlation functions shall
be similar to the one found in the soft N\'eel phase of our model. 
We are presently investigating this point.

\vskip15mm

{\Large\bf Acknowledgements}

\vskip5mm
This work was supported in part by CNPq, FAPERJ and \\ PRONEX No. 66.2002/1998-9.

\vfill
\eject


\begin{thebibliography}{99}


\bibitem{ha} F.D.M.Haldane, {\it Phys. Rev. Lett.} {\bf 50}, 1153 (1983);
F.D.M.Haldane, {\it Phys. Lett.}, {\bf A93} 464 (1983)

\bibitem{chn} S.Chakravarty, B.Halperin and D.Nelson,
{\it Phys. Rev.} {\bf B39}, 2344 (1989)

\bibitem{rs} N.Read and S.Sachdev,
{\it Phys. Rev.} {\bf B42}, 4568 (1990);
A.V.Chubukov, S.Sachdev and J.Ye, {\it Phys. Rev.} {\bf B49}, 11919 (1994);
O.Starykh, {\it Phys. Rev.} {\bf B50}, 16428 (1994);
A.V.Chubukov and O.Starykh, {\it Phys. Rev.} {\bf B52}, 440 (1995);
V.Y.Irkhin and A.A.Katanin, {\it Phys. Rev.} {\bf B55}, 12318 (1997)

\bibitem{cn} Y-C.Chen and A.H.Castro Neto, {\it Phys. Rev.} {\bf B61}, 3772 (2000)

\bibitem{ak} A.P.Kampf, {\it Phys. Rep.} {\bf 249}, 219 (1994)

\bibitem{em1} E.C.Marino, {\it Phys. Rev.} {\bf B61}, 1588 (2000)

\bibitem{ma} E.Manousakis, {\it Rev. Mod. Phys. } {\bf 63}, 1 (1991)

\bibitem{km} R.K\"oberle and E.C.Marino, {\it Phys. Lett.} {\bf B126},
475 (1983)

\bibitem{bp} A.A.Belavin and A.M.Polyakov, {\it JETP Lett.} {\bf 22},
245 (1975)

\bibitem{sw} C.Wan, A.B.Harris and D.Kumar {\it Phys. Rev.} {\bf B48}, 1036 (1993)

\bibitem{ea} S.F.Edwards and P.W.Anderson, {\it J. Phys.} {\bf F5},
965 (1975)

\bibitem{gr} I.S.Gradshteyn and I.M.Ryzhik, {``Table of Integrals, Series
and Products''}, Academic Press, New York, 1980

\bibitem{sg1} K.Binder and A.P.Young, {\it Rev. Mod. Phys.} {\bf 58},
801 (1986)

\bibitem{daf} D.Fiorani et al., {\it Phys. Rev.} {\bf B30}, 2776 (1984)

\bibitem{em2} E.C.Marino, {\it Phys. Lett.} {\bf A263}, 446 (1999); 
E.C.Marino and M.B.Silva Neto, {\it Phys. Rev.} {\bf B}, (2001), in press
(cond-mat/0008186)



\end{thebibliography}
\end{document}